\documentclass[twocolumn,tighten,times]{aastex631}
\def\hawc{{HAWC\nolinebreak[4]\hspace{-.02em}\raisebox{.17ex}{+}}}

\def\m51{{M\nolinebreak[4]\hspace{0.08em}51}}

\newcommand{\um}{$\mu$m}
\def\m51{{M\nolinebreak[4]\hspace{0.08em}51}}

\shorttitle{The magnetic fields of the Antennae galaxies}
\shortauthors{Lopez-Rodriguez, E.}

\begin{document}

\title{Extragalactic magnetism with SOFIA (SALSA Legacy Program). \\
The magnetic fields in the multi-phase interstellar medium of the Antennae galaxies\footnote{SALSA provides a software repository at \url{https://github.com/galmagfields/hawc}, and publicly available data at \url{http://galmagfields.com/}}}

\correspondingauthor{Enrique Lopez-Rodriguez}
\email{elopezrodriguez@stanford.edu}

\author[0000-0001-5357-6538]{Enrique Lopez-Rodriguez}
\affiliation{Kavli Institute for Particle Astrophysics \& Cosmology (KIPAC), Stanford University, Stanford, CA 94305, USA}

\author[0000-0003-3249-4431]{Alejandro S. Borlaff}
\affil{NASA Ames Research Center, Moffett Field, CA 94035, USA}

\author{Rainer Beck}
\affil{Max-Planck-Institut f\"ur Radioastronomie, Auf dem H\"ugel 69, 53121 Bonn, Germany}

\author[0000-0001-8362-4094]{William T. Reach}
\affil{Universities Space Research Association, NASA Ames Research Center, Moffett Field, CA 94035, USA}

\author[0000-0001-8906-7866]{Sui Ann Mao}
\affil{Max-Planck-Institut f\"ur Radioastronomie, Auf dem H\"ugel 69, 53121 Bonn, Germany}

\author[0000-0002-4324-0034]{Evangelia Ntormousi}
\affil{Scuola Normale Superiore, Piazza dei Cavalieri 7, 56126 Pisa, Italy}
\affil{Institute of Astrophysics, Foundation for Research and Technology-Hellas, Vasilika Vouton, GR-70013 Heraklion, Greece}

\author[0000-0002-8831-2038]{Konstantinos Tassis}
\affil{Department of Physics \& ITCP, University of Crete, GR-70013, Heraklion, Greece} 
\affil{Institute of Astrophysics, Foundation for Research and Technology-Hellas, Vasilika Vouton, GR-70013 Heraklion, Greece}

\author[0000-0002-4059-9850]{Sergio Martin-Alvarez}
\affiliation{Kavli Institute for Particle Astrophysics \& Cosmology (KIPAC), Stanford University, Stanford, CA 94305, USA}

\author[0000-0002-7633-3376]{Susan E. Clark}
\affiliation{Department of Physics, Stanford University, Stanford, California 94305, USA}
\affiliation{Kavli Institute for Particle Astrophysics \& Cosmology (KIPAC), Stanford University, Stanford, CA 94305, USA}

\author[0000-0002-5782-9093]{Daniel~A.~Dale}
\affil{Department of Physics \& Astronomy, University of Wyoming, Laramie, WY, USA} 

\author[0000-0001-8931-1152]{Ignacio del Moral-Castro}
\affil{Instituto de Astrof\'isica de Canarias, C/ V\'ia L\'actea, s/n, 38205, La Laguna, Tenerife, Spain}
\affil{Departamento de Astrof\'isica, Universidad de La Laguna, 38206, La Laguna, Tenerife, Spain}

\begin{abstract}
Mergers are thought to be a fundamental channel for galaxy growth, perturbing the gas dynamics and the magnetic fields (B-fields) in the interstellar medium (ISM). However, the mechanisms that amplify and dissipate B-fields during a merger remain unclear. We characterize the morphology of the ordered B-fields in the multi-phase ISM of the closest merger of two spiral galaxies, the Antennae galaxies. We compare the inferred B-fields using $154$ \um~thermal dust and $11$ cm radio synchrotron emission polarimetric observations. We find that the $154$ \um~B-fields are more ordered across the Antennae galaxies than the $11$ cm B-fields. The turbulent-to-ordered $154$ \um~B-field increases at the galaxy cores and star-forming regions. The relic spiral arm has an ordered spiral $154$ \um~B-field, while the $11$ cm B-field is radial. The $154$ \um~B-field may be dominated by turbulent dynamos with high $^{12}$CO(1-0) velocity dispersion driven by star-forming regions, while the $11$ cm B-field is cospatial with high HI velocity dispersion driven by galaxy interaction. This result shows the dissociation between the warm gas mainly disturbed by the merger, and the dense gas still following the dynamics of the relic spiral arm. We find a $\sim8.9$ kpc scale ordered B-field connecting the two galaxies. The base of the tidal tail is cospatial with the HI and $^{12}$CO(1-0) emission and has compressed and/or sheared $154$ \um~and $11$ cm B-fields driven by the merger. We suggest that amplified B-fields, with respect to the rest of the system and other spiral galaxies, may be supporting the gas flow between both galaxies and the tidal tail.
\end{abstract}



\section{Introduction} \label{sec:INT}

The majority of disk galaxies have experienced at least one major merger in their history\,\citep{Hammer2009aap507_1313, Elichemoral2010aap519_A55, Prieto2013}. During major mergers, stars and dark matter can be scattered in extended tidal tails and stellar halos. On the other hand, the gas tends to lose angular momentum and be compressed towards regions of deeper gravitational potential and dense clumps \citep{Teyssier2010}. The compression of highly collisional components of the interstellar medium (ISM), like molecular gas, rapidly enhances star formation in interacting galaxies \citep{Kennicutt1987, Barnes1992}. Furthermore, observational studies have shown that the fraction of molecular gas increases in interacting galaxies \citep{BraineCombes1993, Combes1994, Casasola2004}. This increase is associated with an enhancement of turbulence \citep{Renaud2014}, which generates large amounts of turbulent kinetic energy across the merger.

All observed galaxies host large-scale magnetic fields (B-fields) \citep[e.g.,][]{Beck2019,ELR2022b}. Thus, it is evident that the pre-merger galaxies may host large-scale, ordered B-fields, which can be tangled due to the turbulent kinetic energy generated by the merger. This turbulent kinetic energy can also be available for turbulent dynamo action if the pre-merger B-fields are weak \citep{BE2022}. Note that the typical turbulence coherence length is $<50$--$100$~pc driven by supernova explosions in the ISM \citep[][]{haverkorn+2008apj680_362, bhat+2016mnras461_240,ss21}. On scales larger than the turbulence coherence length, the galactic ISM can be approximated as infinitely conducting. Therefore, the B-field is frozen into the plasma, and since gas is sheared and compressed, the B-field lines will too---the gas flows affect the evolution of B-fields and vice versa. As a consequence, the B-fields can affect galaxy growth and the global structure of galactic disks as shown in magnetohydrodynamical (MHD) simulations \citep{Kotarba2010ApJ7161438K,martinalvarez+2020mnras495_4475,vandevoort+2021mnras501_4888,Whittingham2021}. However, it remains unclear how the B-fields in the multi-phase ISM are amplified and perturbed by mergers. 

\begin{figure*}[ht!]
\includegraphics[width=\textwidth]{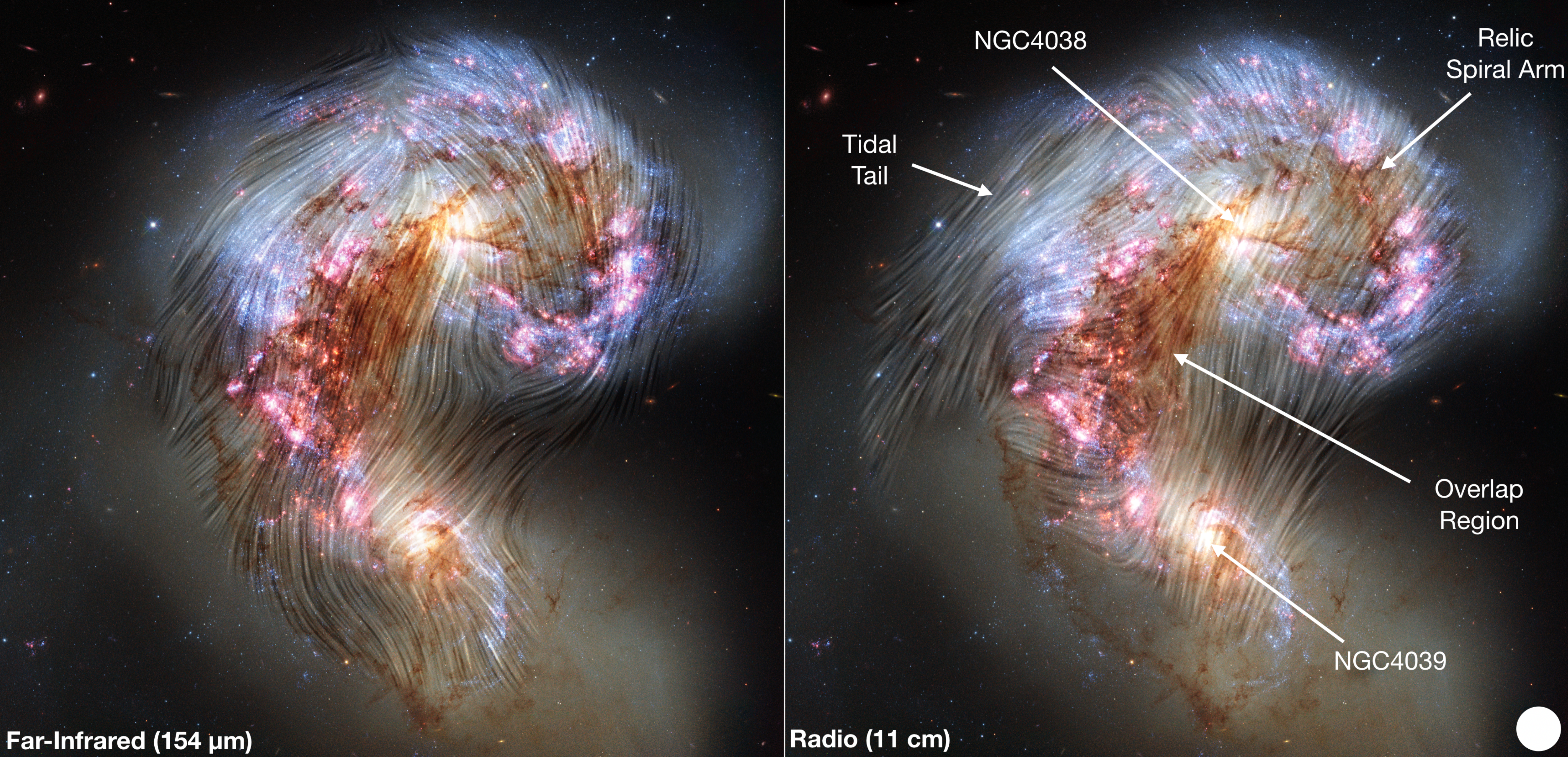}
\caption{B-field orientations in the plane of the sky (streamlines) at wavelengths of $154$~\um\ (left) and $11$~cm \citep[right;][corrected for Faraday rotation]{Basu2017} over a color image of the Antennae galaxies. The color image was computed using \textit{HST} F435W (blue), F550M (green), and a combination of F814W and F658N (red) by \citet{Whitmore2010}. The inferred B-field orientation at $154$~\um\ is shown as streamlines using the Line Integral Convolution \citep{LIC} technique (resample=20). At $154$ \um, we show polarization measurements with $I/\sigma_{I}~\ge~100$, $PI/\sigma_{PI}~\ge~3$, and $P~\le~20$\%, where $\sigma_{I}$ and $\sigma_{PI}$ are the uncertainties in the total and polarized intensities, respectively. At 11 cm, we show polarization measurements with $PI/\sigma_{PI}~\ge~5$. We show the labels of the regions studied in this work and the beam size (white circle) of the $154$ \um~observations.
 \label{fig:fig1}}
\end{figure*}

A comparison between far-infrared (FIR; $154~\mu$m) and radio ($3$ and $6$~cm) polarimetric observations of the \m51\ merging system showed that the two wavelength ranges reveal differences in the morphology of the ordered B-fields \citep{borlaff+2021apj921_128}. The FIR and radio observations were obtained with the Stratospheric Observatory for Infrared Astronomy (SOFIA) and the Very Large Array (VLA), respectively at an angular resolution of $13\arcsec$ (565 pc). The difference in the observed B-field structure arises from the different nature of the ISM associated with these tracers. Radio synchrotron polarization traces the B-field in a volume-filling relativistic medium of the warm and diffuse ISM \citep{Beck2019}. The FIR polarization observations trace magnetically aligned dust grains, which provide the plane-of-sky (POS) B-field morphology, integrated along the line-of-sight (LOS). The measured FIR polarization is weighted by a combination of density, temperature, and grain alignment efficiency along the LOS and within the beam of a dense ($\log_{10}(N_{\rm~HI+H_{2}}~[\rm{cm}^{-2}])=[19.9,22.9]$) and cold ($T_{\rm~d}=[19,48]$ K) component of the ISM \citep[SALSA IV,][]{ELR2022b}. The differences between radio and FIR polarimetric observations allow us to investigate extragalactic B-fields in different ISM phases, something that is uniquely enabled by the Survey for extragALactic magnetiSm with SOFIA (SALSA) Legacy Program \citep{ELR2022b}.

Observations and simulations have shown that the B-field strength increases during mergers, peaking when the cores merge and decreasing in the remnant \citep{Kotarba2010ApJ7161438K,Drzazga2011, Rodenbeck2016,martinalvarez+2020mnras495_4475,Whittingham2021}. One of the first detailed analyses of the B-field structure for galactic mergers was performed on the Antennae galaxies (Arp\,244, NGC\,4038/NGC\,4039). Using $3$ and $6$ cm radio polarimetric observations, \citet{chyzybeck2004} found an average strength of the total B-field of $\simeq~20~\mu$G, two times stronger than in undisturbed spirals, and even $\simeq~30~\mu$G in the interaction region. From Faraday rotation measures in the $7-15$ cm band, \cite{Basu2017} discovered regular B-fields of $\simeq20~\mu$G strength along a 20--kpc sized tidal tail. This B-field is probably a stretched and amplified regular B-field originating from the disk of the progenitor galaxy. These results indicate that compression and tangling driven by the merger activity can amplify the B-fields of the pre-merger galaxies. Furthermore, mergers also enhance star formation activity and hence increase the turbulent kinetic energy of the progenitor galaxies. This turbulent kinetic energy can be available for turbulence-driven dynamos. This emphasizes the importance of understanding which mechanisms (i.e. compression, stretching, tangling, dynamo) are responsible for explaining the magnetization of galaxies since the early Universe, when the merger rate was much higher.

We aim to characterize the B-field morphology in the warm (Radio) and cold (FIR) ISM of the Antennae galaxies, and associate them with the molecular and neutral gas and the star-formation activity. The Antennae galaxies are among the closest  \citep[$D~=~22\pm3$~Mpc; $1\arcsec~=~105$~pc;][using Type~Ia supernova 2007sr]{Schweizer2008} major merging galaxies to the Milky Way, they are highly luminous \citep[$L_{\rm~8-1000\mu m}=7.2\times10^{10}$L$_{\odot}$,][]{Sanders2003}, and have a large angular size (2.6\arcmin~$\sim$~16.4~kpc~in diameter). These characteristics make the Antennae galaxies an excellent target for the High-resolution Airborne Wideband Camera-plus \citep[\hawc,][]{Vaillancourt2007, Dowell2010, Harper2018} onboard SOFIA.

\section{Far-infrared Polarimetric Observations}\label{sec:OBS}

\subsection{Observations, data reduction, and complementary data}\label{sec:OBS}

\begin{figure*}[ht!]
\centering
\includegraphics[scale=0.52]{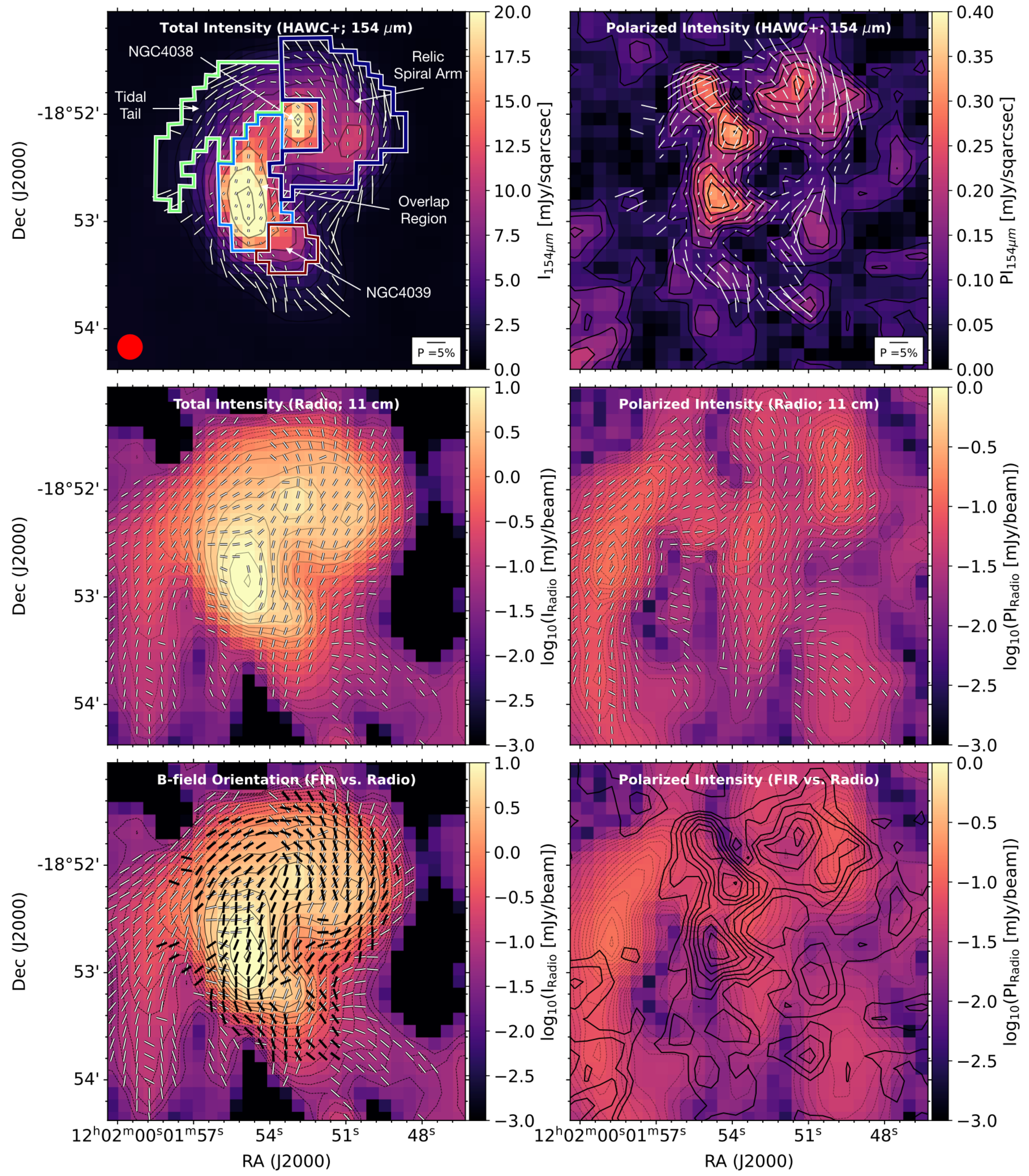}
\caption{Total and polarized intensities of the Antennae galaxies at wavelengths of $154$~\um\ and $11$~cm. 
\textit{Top:} $I_{\rm~154\mu m}$ (left; colorscale) with overlaid contours starting at $100\sigma_{\rm~I_{154~\mu m}}$ and increasing in steps of $2^{n}\times\sigma_{\rm~I_{154~\mu m}}$, where $n=6.7, 7.2, 7.7, \dots$ and $\sigma_{\rm~I_{154~\mu m}}~=~0.01$~mJy/sqarcsec. $PI_{\rm~154\mu m}$ (right; colorscale) with overlaid contours at $3\sigma_{\rm~PI_{154~\mu m}}$ and increasing in steps of $n\times\sigma$, where $n=3, 3.5, 4, \dots$ and $\sigma_{\rm~PI_{154~\mu m}}~=~0.07$~mJy/sqarcsec. For both figures, the  $154$~\um\ B-field orientations (white lines) with their lengths proportional to $P$ are shown. A legend of 5\% polarization, the beam size (red circle), and several physical structures with their outlined mask (Appendix \ref{App:ArchivalData}) are shown. 
\textit{Middle:} $I_{\rm~radio}$ (left; colorscale) with overlaid contours in $\log_{10}$ scale starting at $4\sigma_{\rm~I_{Radio}}$ and increasing in steps of $2^{n}\times\sigma_{\rm~I_{Radio}}$, where $n=2, 2.5, 3, \dots$ and $\sigma_{\rm I_{Radio}} = 0.01$ mJy/beam. $PI_{\rm radio}$ (right; colorscale) with overlaid contours in $\log_{10}$ scale starting at $5\sigma_{\rm PI_{Radio}}$ and increasing in steps of $2^{n}\times\sigma_{\rm PI_{Radio}}$, where $n=5, 7, 9, \dots$ and $\sigma_{\rm PI_{Radio}} = 0.004$ mJy/beam. The lengths of the lines are set to unity to show the B-field orientations.
\textit{Bottom:} Overlaid comparison of the B-field orientation (left) and $PI$ (right) at $154$~\um\ (black lines and contours) and $11$~cm (white lines and grey contours). For both figures, $PI$ is displayed in colorscale as described above.
 \label{fig:fig2}}
\end{figure*}

The Antennae galaxies were observed as part of SALSA (ID: 08\_0012; PI: Lopez-Rodriguez, E. \& Mao, S. A.) on 2022-06-02 (flight ID: F880) and 2022-06-11 (F886) from Palmdale, CA, and 2022-06-21 (F889) and 2022-06-27 (F891) during the SOFIA deployment in Christchurch, New Zealand. The galaxies were only observed with HAWC+ at $154$ \um\ with a pixel scale of $6\farcs9$ and beam size (FWHM) of $13$\farcs$6$. We used the on-the-fly-mapping polarimetric mode with a scan rate of $200$\arcsec sec$^{-1}$, scan amplitude of $100\arcsec \times 100\arcsec$, and scan on-source time of $120$s per halfwave plate position angle.

The observations were reduced using the same pipeline scheme as described in SALSA III \citep{ELR2022a}.  The final data products contain the Stokes $IQU$, polarization fraction ($P$), position angle of polarization ($PA$), polarized intensity ($PI$), and their uncertainties, with a pixel scale equal to the detector pixel scale, $6\farcs9$, at $154$ \um. The polarization fraction was debiased \citep{Serkowski1962,WK1974} and corrected for instrumental polarization. The final data product has a total on-source time of $4.8$h with a total execution time of $5.1$h. The final Stokes $I$ was spatially correlated with the \textit{Herschel} image at $160$~\um\ to correct for the WCS, where small corrections of $1-2$ pixels were required. The Stokes $QU$ and their uncertainties were also corrected using the same offsets. The flux calibration uncertainty is estimated to be $8$\%. We also included the intrinsic polarization uncertainty of $0.2$\% and angular uncertainty of $3^{\circ}$ of HAWC+ \citep{Harper2018, ELR2022a}. Note that our analysis is performed with polarization measurements having $P/\sigma_{P} \ge3$, which ensures an uncertainty in the angle of polarization of $\le9.6^{\circ}$. Individual uncertainties are accounted for in the analysis of angular dispersion throughout this work.

We use a set of archival data to support the analysis of the HAWC+ observations (Appendix \ref{App:ArchivalData}). We use $11$ cm radio polarimetric observations from the VLA with an angular resolution of $11\arcsec\times9\arcsec$  \citep{Basu2017}. These radio polarimetric observations, corrected for Faraday rotation, are used to trace the B-fields in the warm and diffuse ISM. We use the \textit{HST}/WFC/ACS F658N image covering the H$_{\alpha}$+NII emission lines \citep{Whitmore2010} as a proxy for the ongoing star formation in the galaxies. We use the HI emission line observations with an angular resolution of $11.41\arcsec\times7.48\arcsec$  \citep{Hibbard2001} and the $^{12}$CO(1-0) line observations with an angular resolution of $15.2\arcsec\times7.8\arcsec$. These observations are used as a proxy for the neutral and molecular gas dynamics in the galaxy, respectively. We also use the X-ray \textit{Chandra}/ACIS images with an angular resolution of $\sim8$\arcsec\ covering the full energy range of $0.1-10.0$ keV \citep{Fabbiano2001} as a proxy for the hot ISM associated with the active star-forming regions located in the warm ISM traced by the H$_{\alpha}$ emission line. In addition, we compute the column density, $N_{\rm HI+H_{2}}$, and dust temperature, $T_{\rm d}$, maps across the Antennae galaxies using $70-250$ \um~\textit{Herschel} observations. Data were smoothed and projected to the HAWC+ pixel grid as described in Appendix \ref{App:ArchivalData}.

\subsection{Results}\label{sec:RES}

\begin{figure*}[ht!]
\includegraphics[width=\textwidth]{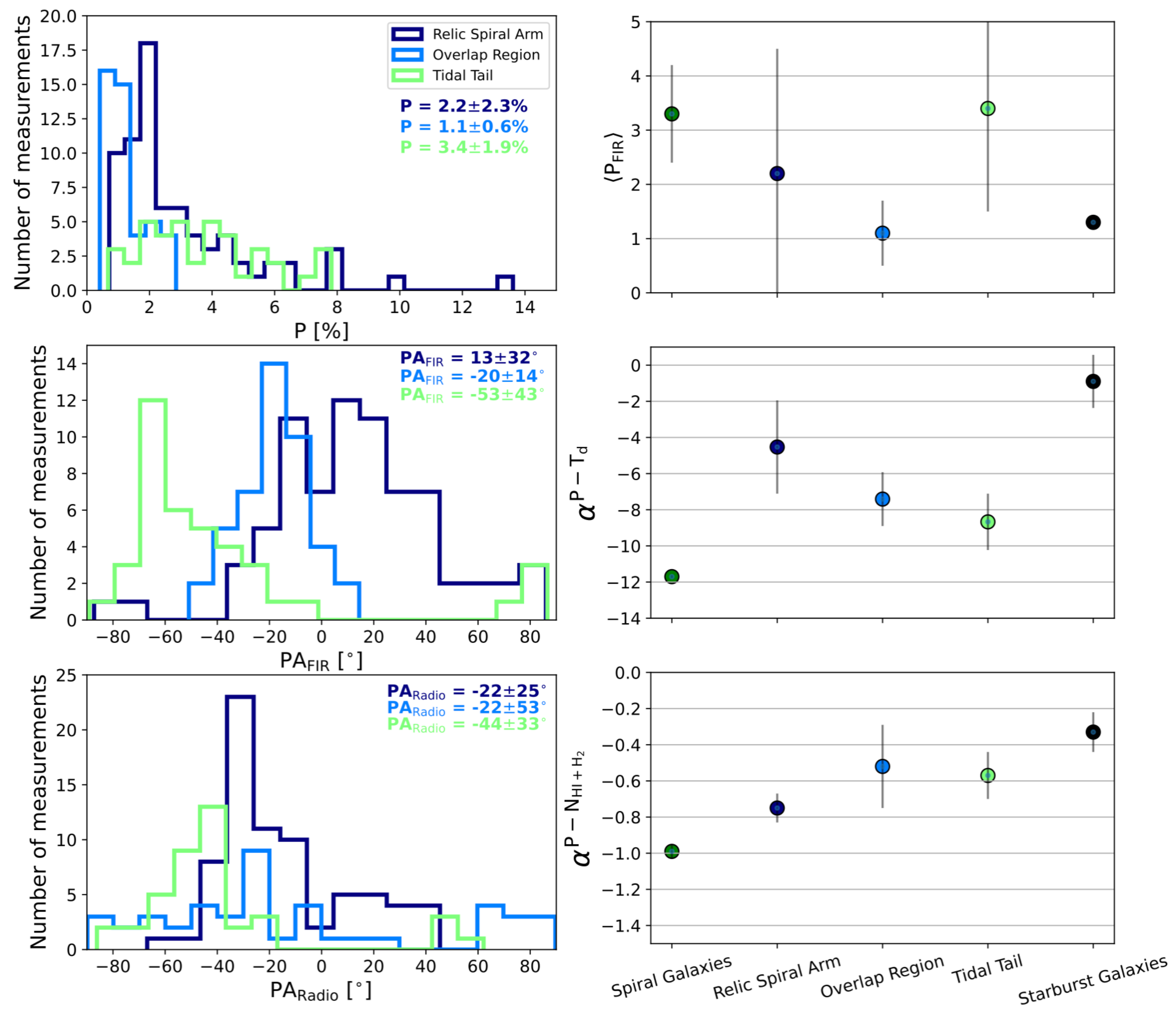}
\caption{Comparison of the polarization measurements between radio and FIR wavelengths for the different regions studied.
\textit{Left:} The histograms of $P$ (top) in $0.5$\% bins and the B-field orientation in bins of $10^{\circ}$ at $154$~\um\ (middle) and $11$~cm (bottom) are shown. The median and $1\sigma$ uncertainties are shown. 
\textit{Right:} Median $154$~\um\ $P$ (top), the slopes (Figure \ref{fig:fig4}) of the $P-T_{\rm d}$ (middle) and $P-N_{\rm HI+H_{2}}$ (bottom) of the Antennae's regions, and for spiral (M51, M83, NGC~3627, NGC~4736, NGC~6946, NGC~7331) and starburst (M82, NGC~253, NGC~2146) galaxies at $154$ \um~taken from \citet{ELR2022b} are shown.
 \label{fig:fig3}}
\end{figure*}

The most remarkable result is the large-scale ordered B-field structures across the $\sim13$~kpc diameter of the Antennae galaxies (Fig.\,\ref{fig:fig1}). We label the several features in the Antennae galaxies as shown by \citet{chyzybeck2004}. We measure a polarization fraction across the galaxies within the range of $[0,8]$\% (Figs.\,\ref{fig:fig2} and \ref{fig:fig3}), with some outliers within $[10,13]$\% located in the most northern region of the relic spiral arm and the southern region from NGC~4039. We estimate the median polarization fraction and B-field orientation for several components across the Antennae galaxies using the mask described in Appendix \ref{App:ArchivalData}. Note that the quoted uncertainties represent the dispersion of all the individual measurements within each region, and should not be interpreted as the uncertainty of the polarization measurement. 

The overlap region has an ordered B-field scale of $\sim8.9$~kpc at a fairly constant orientation of $-20\pm14^{\circ}$ (Figs.\,\ref{fig:fig2} and \ref{fig:fig3}). This B-field structure connects the core of NGC~4038 (northern core) with the northern region of the core of NGC~4039 (southern core). The maximum polarization fraction is $\sim2$\% in the middle and drops to $\le0.4$\% at the northern and southern edges. 

The relic spiral arm has a tightly wrapped spiral B-field morphology. The median polarization fraction is $2.2\pm0.3$\%. An unpolarized ($<0.4$\%) region is identified at the start of the relic spiral arm. This is the region where the arm twists from a southwest orientation starting from NGC~4038 to a north orientation. The B-field continues wrapping around the core of NGC~4038 to a northeast orientation. 

Our observations are sensitive to the first $\sim3.2$~kpcs ($\sim0.5\arcmin$) of the tidal tail. This region has the highest measured median polarization fraction of $3.4\pm1.9$\%, and has a highly ordered B-field. The B-field orientation, $\sim-53^{\circ}$, is more open than the spiral B-field in the relic spiral arm.  We find a valley in the polarization fraction in the interface between the end of the relic spiral arm and the beginning of the tidal tail. 

NGC~4039 and the southern region from the core of NGC~4038 are unpolarized ($<0.4$\%). NGC~4038 has a polarization of $\sim0.7$\%.

We compute the integrated polarization fraction and B-field orientation of the Antennae galaxies following the same procedure as described in SALSA IV \citep[][section 4.1]{ELR2022b} to account for the vector quantity of the polarization measurements. The integrated polarization fraction results in a weakly polarized, $\langle P^{\rm int}_{\rm 154~\mu m} \rangle = 0.8\pm0.1$\%, system with a B-field orientation,  $\langle PA^{\rm int}_{\rm 154 \mu m} \rangle = -10\pm5^{\circ}$, roughly parallel to the B-field orientation, $\sim-20^{\circ}$, of the overlap region. Figure \ref{fig:fig2} shows that approximately half of the total polarized emission, $0.97\pm0.03$~Jy, arises from the overlap region, $0.44\pm0.04$~Jy, and less than half from the relic spiral arm, $0.36\pm0.05$~Jy. Both of these regions have different B-field orientations, which end up with a net angle of $-10^{\circ}$ in the integrated B-field orientation due to the vector properties of the polarization.

\section{The Magnetic fields at FIR and radio wavelengths}\label{subsec:FIRvsRadio}

\begin{figure*}[ht!]
\centering
\includegraphics[width=\textwidth]{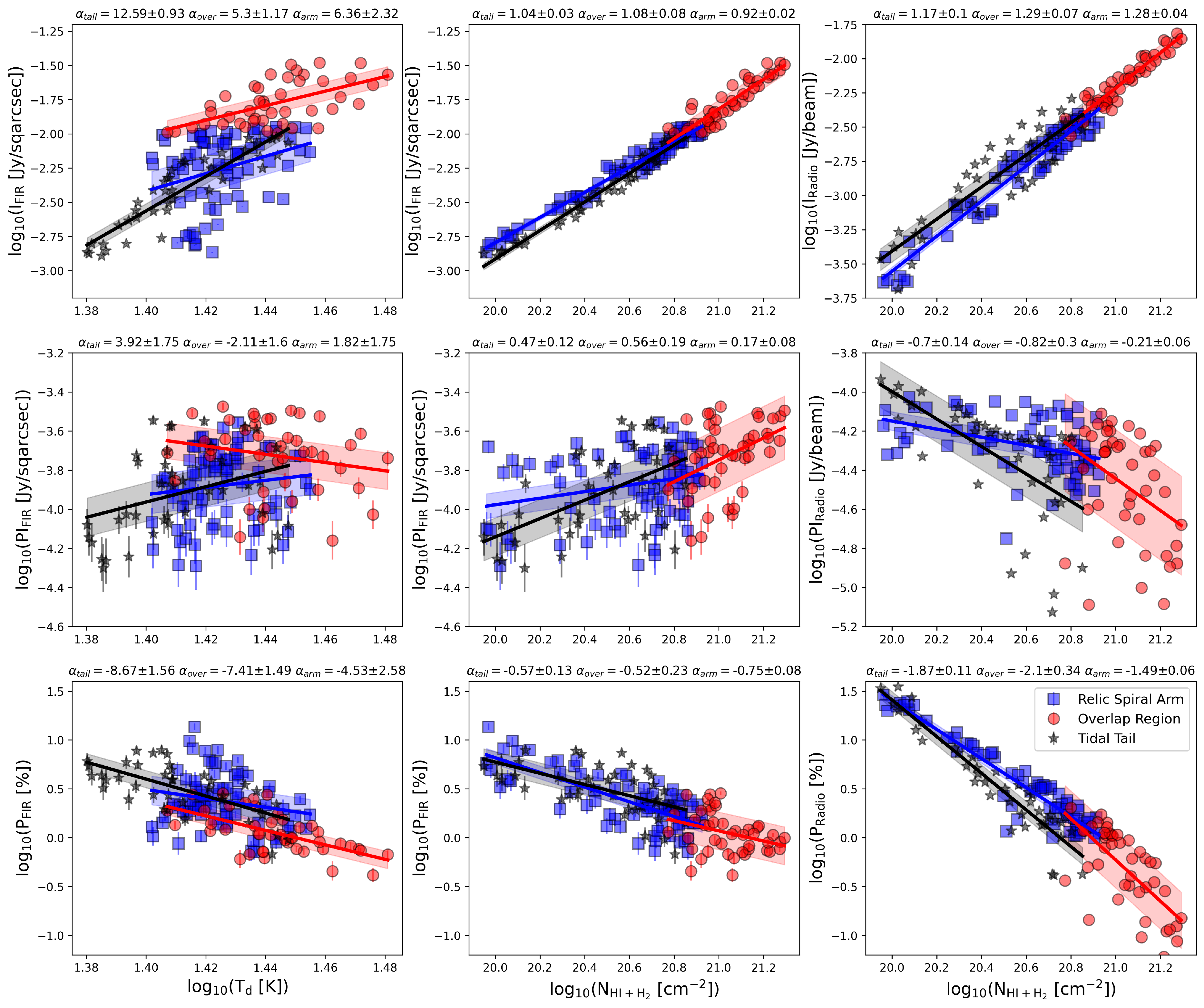}
\caption{Polarization measurements vs. dust temperature and column density for three regions: relic spiral arm, overlap region, and tidal tail. The $I$ (top), $PI$ (middle), and $P$ (bottom) vs. $T_{\rm d}$ (left) and $N_{HI+H_{2}}$ (middle and right) at $154$~\um\ (left and middle) and $11$~cm (right) are shown. The best fit model and the slopes with their uncertainties for each region are shown in each panel.
 \label{fig:fig4}}
\end{figure*}

\subsection{Comparison of the magnetic field morphologies}\label{subsec:PIplots}

The $11$~cm B-field (corrected for Faraday rotation) orientations show several different morphological structures to the $154$~\um\ B-fields (Fig. \ref{fig:fig1} and \ref{fig:fig2}). We compare the same LOS radio polarization measurements to those at $154$ \um, both at a common resolution and on the same pixel grid (Appendix \ref{App:ArchivalData}). The radio observations showed a large-scale, $\sim20$~kpc, coherent B-field structure following the tidal tail \citep{Basu2017}. Our FIR observations unequivocally trace the same B-field orientation within the first $\sim3.2$~kpc-scale of the tidal tail. Note that the radio polarized flux also has a polarization valley at the end of the relic spiral arm and the start of the tidal tail. We estimate a median radio B-field orientation of $-44\pm33^{\circ}$ in the tidal tail, with a lower angular dispersion than at $154$~\um. 

The largest differences in the B-field orientations between FIR and radio are found at the location of the relic spiral arm and the east side of the overlap region.  At the location of the relic spiral arm, the $11$~cm observations show an almost radial B-field, $-22\pm25^{\circ}$, while the B-field is tightly wrapped at $154~\mu$m. At both $3.6$ and $6.2$~cm radio wavelengths, a spiral B-field with a large pitch angle is observed in the relic spiral arm \citep{chyzybeck2004}. Probably these observations are less affected by Faraday depolarization than at $11$~cm and trace the B-fields closer to the cold gas in the relic spiral arm. 

We estimate a median radio B-field orientation of $-22\pm53^{\circ}$ in the overlap region, which has larger angular dispersion than at $154$~\um. At radio wavelengths, the B-field orientation shows a twist from the core of NGC~4038 to the tidal tail in the eastern region of the overlap region. The $11$~cm B-field orientations differ up to $90^{\circ}$ from that traced at $154~\mu$m. On the western side, the radio B-field orientation is parallel to that traced at $154~\mu$m connecting both galaxies.

We compute the integrated $11$~cm B-field orientation, as described in Section \ref{sec:RES}, to be $\langle PA^{\rm int}_{\rm Radio} \rangle = -29\pm7^{\circ}$. The angular dispersion in both FIR and radio are similar, indicating a comparably complex environment. The median integrated B-field orientations differ at both wavelengths. This result is mainly due to the differences in the polarized flux emission across the galaxies (Fig. \ref{fig:fig2}--bottom right). The radio polarized emission is mainly arising from the outskirts of the relic spiral arm and the western side of the overlap region.

\subsection{Relative contribution of the ordered and random magnetic fields}\label{subsec:PIplots}

We compare the $154$~\um\ and $11$~cm total fluxes and polarization measurements with the column density, $N_{\rm HI+H_{2}}$ (Fig. \ref{fig:fig4}). The $N_{\rm HI+H_{2}}$ were estimated using $70-250$ \um\ \textit{Herschel} observations (Appendix \ref{App:ArchivalData}). The $P-N_{\rm HI+H_{2}}$ relationship is sensitive to the isotropic random and/or tangled B-fields along the LOS within the beam of the observations. Thus, steeper negative slopes are expected with increasing turbulence and/or tangled B-fields at FIR wavelengths. A negative slope is also expected when decreasing anisotropic random B-fields \citep[][]{Beck2019,ELR2022b}, because the net polarization of isotropic random B-fields summed over the beam will approach zero, resulting in null measured polarization. At radio wavelengths, the strength of the total B-field (dominated by isotropic random B-fields) increases with gas density, and the polarized flux increases as anisotropic turbulent B-fields increases \citep{Beck2019}.

We find that the $154$~\um\ and $11$~cm total intensities increase with $N_{\rm HI+H_{2}}$. We estimate steeper positive slopes at radio than FIR wavelengths for all regions, which indicates a large contribution of isotropic random B-fields at radio wavelengths. The $PI-N_{HI+H_{2}}$ plots show opposite trends for FIR (positive) and radio (negative) wavelengths. Since the polarization fraction is given by $P=PI/I$, the decrease of $P$ with $N_{\rm HI+H_{2}}$ is steeper at $11$~cm than at $154$ \um. At $154$ \um, the positive trends in $PI-N_{\rm HI+H_{2}}$ may be explained due to an increase of anisotropic, random B-fields by compression and/or shearing in the cold gas. This trend is in agreement with those previously measured in galactic outflows in starburst galaxies and galactic shock-driven regions in spiral galaxies \citep[Fig.\,5-left in SALSA IV;][]{ELR2022b}, but would be opposite to the trend seen in radio.  At $11$~cm, the negative trends in $PI-N_{\rm HI+H_{2}}$ may be explained by an increase of isotropic random B-fields in the warm gas. The isotropic and/or random B-fields decrease the net polarization within the beam of the radio observations \citep{Beck2019}. Note that this trend is opposite to that found in M51 \citep[fig. 14,][]{borlaff+2021apj921_128}. For M51, the $PI_{\rm radio}-N_{\rm HI+H_{2}}$ plot shows a positive trend attributed to an increase of anisotropic turbulent fields in the disk. These results show that the cold and dense ISM is embedded in a relatively more ordered B-field than the warm and diffuse ISM across the Antennae galaxies.

For the relic spiral arm, the $PI-N_{\rm HI+H_{2}}$ plots at FIR show larger dispersion than at radio. For both the overlap region and the tidal tail, the negative (positive) $PI-N_{\rm HI+H_{2}}$ slopes in radio (FIR) are very steep but also show large dispersion. Note that we are only sensitive to the start of the tidal tail, and that the $PI_{\rm radio}$ increases as the distance increases from this region along the tidal tail. The slope in the relic spiral arm is also flatter than in the overlap region and the tidal tail---the overlap region has higher dust temperature (T$_{{\rm d}}$) with similar N$_{\rm HI+H_{2}}$ than the tidal tail. These results may indicate that different physical conditions may be present in each region, as discussed in Section \ref{subsec:ISM}.

\section{Discussion and final remarks}\label{sec:DIS}

\begin{figure*}[ht!]
\centering
\includegraphics[width=\textwidth]{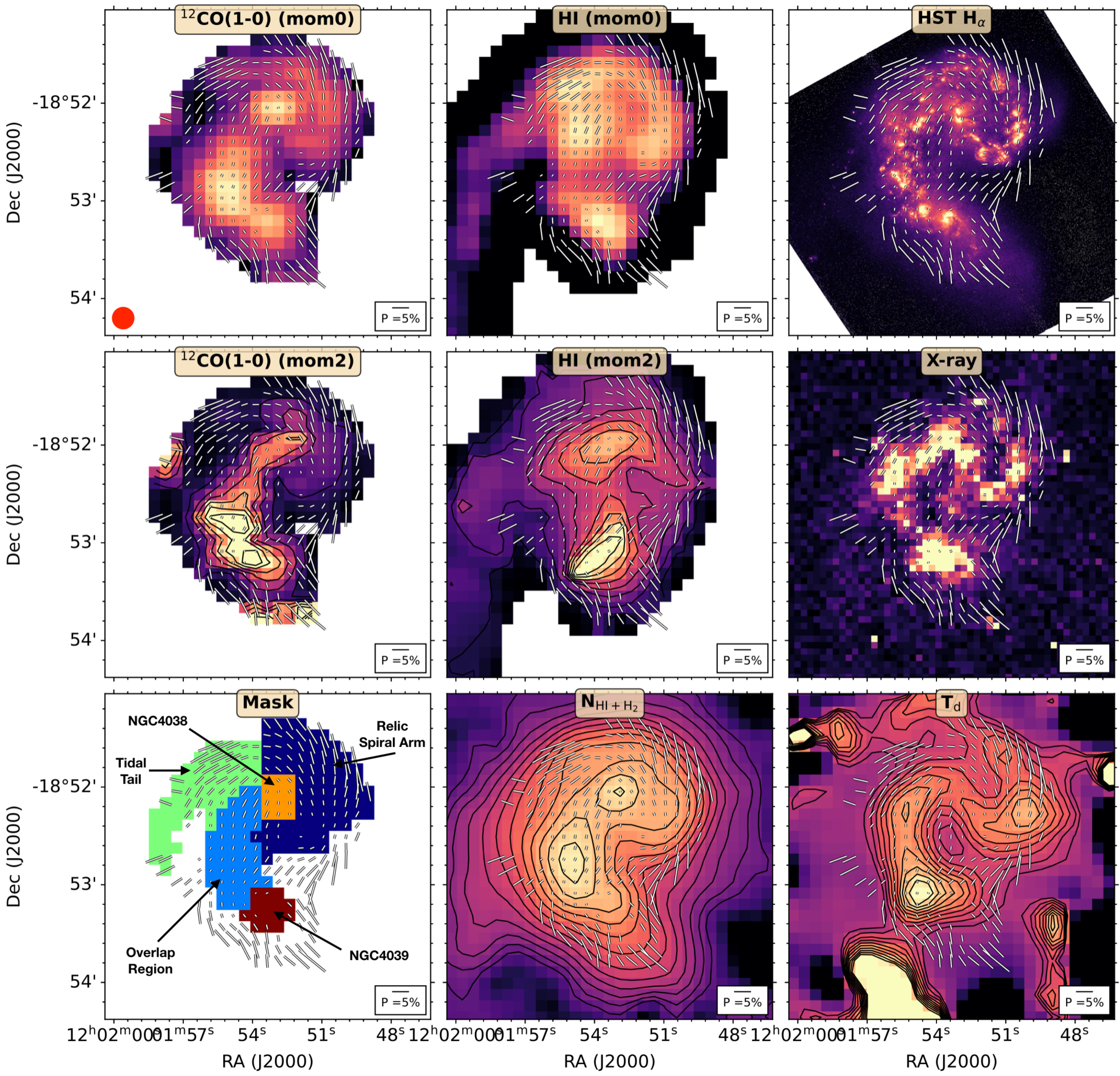}
\caption{Comparison of the B-field orientation at $154$ \um\ with the multi-phase ISM. The B-field orientations at $154$ \um\  (white lines) over the $^{12}$CO(1-0) (top-left) and HI (top-middle) emission lines, H$_{\alpha}$ (top-right), $^{12}$CO(1-0) and (middle-left) and HI velocity dispersion (middle-middle), X-ray (middle-right), mask (bottom-left), $N_{\rm HI+H_{2}}$ (bottom-middle), and $T_{d}$ (bottom-right) are shown. We display the same B-field orientation as shown in Figure \ref{fig:fig2}. The contours in $N_{\rm HI+H_{2}}$ start at $\log_{10}(N_{\rm HI+H_{2}} [\rm{cm}^{-2}]) = 18$ and increase in steps of $0.2$. The contours in $T_{\rm d}$ start at $20$ K and increase in steps of $0.5$ K. The contours in the $^{12}$CO(1-0) and HI velocity dispersion start at $10$ km s$^{-1}$ and increase in steps of $10$ km s$^{-1}$. The beam size (red circle) of the $154$ \um~observations is shown.
 \label{fig:fig5}}
\end{figure*}

\subsection{Magnetic fields in the multi-phase interstellar medium}\label{subsec:ISM}

We study the B-field morphological correspondence with several tracers of the multi-phase ISM (Fig. \ref{fig:fig5}). In the tidal tail, the $PI_{\rm FIR}$ and $PI_{\rm radio}$ are correlated with the HI and $^{12}$CO(1-0) total integrated emission lines (i.e. moment 0), and anti-correlated with their velocity dispersion (i.e. moment 2), and the H$_{\alpha}$ and X-ray emission. We use the velocity dispersion as a proxy for the turbulent kinetic energy of the gas. We conclude that the increase of $PI_{\rm FIR}$ and $PI_{\rm radio}$ are due to a decrease in the neutral and molecular gas turbulence in the tidal tail. The $20$--kpc tidal tail has highly coherent and strong regular B-fields with strengths of $\sim8~\mu$G \citep{Basu2017}. These results indicate that the B-fields in both the warm and cold ISM may be compressed, sheared, and/or stretched along the tidal tail due to the gravitational forces driven by the merger. Note that our steep negative slope in $P-N_{\rm HI+H_{2}}$ indicates an increase of isotropic random B-fields towards higher $N_{\rm HI+H_{2}}$ in the first $\sim3.2$ kpc the of the tidal tail. These results indicate that the compressed B-fields may increase radially from the start of the tidal tail to its main structure.

In the overlap region, we find that the $PI_{\rm FIR}$ is cospatial, and anti-correlated, with the $^{12}$CO(1-0) velocity dispersion, and correlated with the H$_{\alpha}$ and X-ray emission (proxies of star formation activity). These results indicate that turbulent dynamo is dominant and driven by turbulence in star-forming regions. Interestingly, $PI_{\rm radio}$ is co-spatial, and anti-correlated, with the HI velocity dispersion. The B-field orientation is almost north-south and connects both cores. This region delineates the western side of the overlap region, which lacks star-formation activity. This region has the strongest B-field strength, $\sim30~\mu$G \citep{chyzybeck2004}. We conclude that the strong B-fields ($\sim20~\mu$G) may be quenching star formation on the western side of the overlap region and may be driving the gas flows between both galaxies.

In the relic spiral arm, $I$ and $PI$ at FIR and radio are correlated. However, $PI_{\rm radio}$ is mostly located in the outskirts of the relic spiral arm (Fig. \ref{fig:fig2}-bottom right). The $PI_{\rm FIR}$ is co-spatial with the HI and $^{12}$CO(1-0) emission, and the H$_{\alpha}$ and X-ray emission. The $PI_{\rm FIR}$ is anti-correlated with the HI and $^{12}$CO(1-0) velocity dispersion. Note that there is lower H$_{\alpha}$ and X-ray emission, as well as lower HI and $^{12}$CO(1-0) velocity dispersion, in the relic spiral arm than in the overlap region. This result indicates that some areas of the relic spiral arm have lower star-formation activity than in the overlap region. Indeed, \citet{Zhang2010} estimated an averaged star-formation rate (SFR) of $\ge10$ and $\ge1$ M$_{\odot}$ yr$^{-1}$ in the overlap region and the northern region of the relic spiral arm, respectively. The southern region of the relic spiral arm has $SFR\ge10$ M$_{\odot}$ yr$^{-1}$, which is colocated with the unpolarized region at FIR and radio wavelengths. We find that the neutral and molecular gas have a tight spiral B-field in the relic spiral arm, while the warm and diffuse gas has an almost entirely radial B-field from the core of NGC~4038 with no clear morphological correspondence with the neutral and molecular gas. Thus, the B-fields in the warm and the cold gas are dissociated or the radio and FIR are tracing different components of the B-field along the LOS.

We conclude that FIR observations may be tracing the B-fields in the relic spiral arm that have not been strongly perturbed by the merger. These B-fields are highly ordered with low turbulence, which may be decreasing the star-formation activity along the northern region of the relic spiral arm. Furthermore, the B-fields in the warm gas may be dissociated from the gas dynamics of the relic spiral arm. The radio B-fields may be following the extraplanar gas dynamics on the near side of the relic spiral arm. Note that the B-fields at $3.6$ and $6.2$ cm have a more open spiral B-field pattern than at FIR \citep{chyzybeck2004}. These radio wavelengths may be more sensitive to the B-field near the plane of the galaxy.

\subsection{Comparison with other galaxies}\label{subsec:Comp}

We compare (Fig. \ref{fig:fig3}) our measurements in the Antennae galaxies with previous results of spiral and starburst galaxies \citep[SALSA IV,][]{ELR2022b}. The spiral galaxies were found to have large-scale ordered B-fields with their peak of polarization co-located with the interarms and decreasing within the star-forming regions. The starburst galaxies were found to have large-scale ordered B-fields along the galactic outflow. Note that a dominant anisotropic turbulent (random) B-field generates a higher polarization fraction than a dominant isotropic turbulent B-field. These results showed that starburst galaxies have a lower polarization fraction than spiral galaxies. 

For all the above regions, we find that the slopes of the $P-N_{\rm HI+H_{2}}$ plots are between the spiral and the starburst galaxies (Fig. \ref{fig:fig3}). Our results suggest that the relative contribution of the compressed and/or sheared B-fields increases from spiral galaxies to starburst galaxies. The fact that the measured polarization fraction in the overlap region is consistent with that from the starburst galaxies indicates that the overlap region is dominated by highly turbulent B-fields that are compressed, sheared, and/or stretched. For the overlap region, compression and shearing are driven by the merger. The tidal tail has a larger relative contribution of compressed and/or sheared B-fields than the relic spiral arm. The relic spiral arm seems to have a relatively more ordered B-field than spiral galaxies. 

Since the $P-T_{\rm d}$ is sensitive to the radiation field from star formation regions in the ISM of the galaxy, the flatter slope in starburst galaxies indicates that the dust grain alignment efficiency is higher, or are less tangled B-fields along the LOS in the galactic outflows than in the disk of spiral galaxies \citep[SALSA IV,][]{ELR2022b}. The relic spiral arm region has the flattest $P-T_{\rm d}$ slope pointing to a less tangled or turbulent, or higher dust grain alignment efficiency in the cold gas than in the overlap region and tidal tail.  

In this letter, we have reported the detection of ordered B-fields embedded in the cold and dense ISM in the Antennae galaxies using 154 \um\ polarimetric SOFIA/HAWC+ observations. This result is remarkable even though these galaxies are highly perturbed due to the merger. Our observations showed that dust grains aligned with respect to the local B-field are present in the intergalactic medium of mergers. We identified the B-field configurations by measuring the dependency of the polarization fraction with the ISM physical conditions of the merger. The integrated polarization measurements of the merger can be potentially helpful for unresolved polarimetric observations of mergers and/or high redshift galaxies.


\begin{acknowledgments}
The authors thank the reviewer for their helpful comments which improved the quality of the manuscript. E.L.-R thanks Daniel Espada for making the ALMA observations available and guiding the ALMA data processing. We thank Aritra Basu for sharing the radio polarimetric observations. KT has received funding from the European Research Council (ERC) under the European Unions Horizon 2020 research and innovation programme under grant agreement No. 771282. Based on observations made with the NASA/DLR Stratospheric Observatory for Infrared Astronomy (SOFIA) under the  08\_0012 Program. SOFIA is jointly operated by the Universities Space Research Association, Inc. (USRA), under NASA contract NNA17BF53C, and the Deutsches SOFIA Institut (DSI) under DLR contract 50 OK 0901 to the University of Stuttgart. This research made use of astrodendro, a Python package to compute dendrograms of Astronomical data (\url{http://www.dendrograms.org/}).
\end{acknowledgments}


\appendix

\section{Supplementary material}\label{App:ArchivalData}

We use the 2--4\,GHz (central frequency of $2.8$ GHz or $11$ cm wavelength) radio polarimetric observations from the VLA with an angular resolution of $11\arcsec\times9\arcsec$. and a PA $=29^{\circ}$ \citep{Basu2017}.  These images were smoothed using a 2D Gaussian profile with a width equal to the angular resolution, $13\farcs6$, of the HAWC+ observations. Then, the smoothed images were reprojected to the field-of-view (FOV) and the pixelscale, $6\farcs9$, of the HAWC+ observations. In addition, we use the HI emission line observations with an angular resolution of $11.41\arcsec\times7.48\arcsec$ and a PA $=75.63^{\circ}$ taken with the VLA \citep{Hibbard2001}. We use the $^{12}$CO(1-0) line observations with an angular resolution of $15.2\arcsec\times7.8\arcsec$ and a PA $=84^{\circ}$ taken from the Atacama Large (sub-)Mm Array (ALMA) Archive (Project ID: 2017.1.00771.S). The moments 0 and 2 were estimated using the $2500-3050$ channels and a sigma clip of $1.5\sigma$, where $\sigma = 0.091$ Jy beam$^{-1}$. For both $^{12}$CO(1-0) and HI observations, the moments 0 and 2 were smoothed using a 2D Gaussian profile with a width equal to the angular resolution, $13\farcs6$, of the HAWC+ observations. Then, the smoothed images were reprojected to the field-of-view (FOV) and the pixelscale, $6\farcs9$, of the HAWC+ observations. 

We estimated the column density, $N_{\rm HI+H_{2}}$, and dust temperature, $T_{\rm d}$, across the Antennae galaxies. Specifically, we used the $70-250$ \textit{Herschel} PACS and SPIRE observations \citep{Klaas2010}. These images were smoothed using a 2D Gaussian profile with a width equal to the angular resolution, $13\farcs6$, of the HAWC+ observations. Then, the smoothed images were reprojected to the FOV and the pixelscale, $6\farcs9$, of the HAWC+ observations. This approach ensures that images at all wavelengths have the same pixel scale and array dimensions. However, note that the $250$ \um~SPIRE observations have an angular resolution of 17\farcs6, which implies that the $N_{\rm HI+H_{2}}$ and $T_{\rm d}$ maps are slightly oversampled by $4\arcsec$ (below a single detector pixel of HAWC+).  Then, for every pixel we fit an emissivity modified blackbody function with a characteristic dust temperature, $T_{\rm d}$, and emissivity index, $\beta$. \citet{Karl2013} found that multiple modified blackbody functions are required to fit the spectral energy distribution within the $70-500$ \um\ wavelength range. The emissivity indices range from $1.6$ with a $T_{\rm d} = 50$ K and $2.0$ with a $T_{\rm d}= 23$ K. We took an intermediate value of $\beta=1.8$ fixed across the galaxies.  We derived the molecular hydrogen optical depth as $N_{\rm HI+H_{2}} = \tau_{250}/ (\kappa_{250} m_{\rm H}$), where  $\tau_{250}$ is the optical depth at $250$ \um, the dust opacity $\kappa_{250} = 0.1$ cm$^{2}$ g$^{-1}$ at $250$ \um, and the mean molecular weight per hydrogen atom $\mu = 2.8$ \citep{H1983}. The dust temperature and column density values ranges are estimated to be $T_{\rm d} = [22.7,30.3]$ K and  $\log_{10}(N_{\rm HI+H_{2}} [\rm{cm^{-2}}]) = [19.9,21.3]$ in agreement with \citet{Karl2013}.

We used the Python program \textsc{astrodendro}\footnote{\url{http://www.dendrograms.org/}} to identify several regions (Fig. \ref{fig:fig1}) used in the analysis of this work \citep{Rosolowsky2008}. The algorithm identifies local maxima above a defined minimum intensity level with a certain minimum size. We selected each local maximum to be at least two times the beam size, $13\farcs6$ of the HAWC+ observations and the minimum intensity to be $5\sigma$, where $\sigma$ is the standard deviation of the pixel-to-pixel variation in a region of the array without signal from the object. Unfortunately, we found that a single image (i.e. radio, FIR, gas tracer) is not able to separate each region. This is expected because each of these images is sensitive to different phases of the ISM, where each region has different physical properties. We used the smoothed and reprojected images as described above. The relic spiral arm was identified using the HI total intensity for the bulk of the emission and the radio polarized flux for the outskirts regions. The tidal tail was identified using the radio polarized emission. The overlap region was identified using the FIR total emission. NGC~4038's core was identified using the FIR total emission, and the NGC~4039's core using the HI total emission. The final mask is shown in Figure \ref{fig:fig5}. Note that the surrounding regions of NGC~4039 are missing. We did not find a specific tracer for this region, and it was rejected from the study. Further analysis is required to refine the mask and identify the physical structure surrounding NGC~4039.

%

\vspace{5mm}
\facilities{SOFIA (HAWC+), ALMA, \textit{Herschel} (PACS, SPIRE), \textit{HST} (ACS), \textit{Chandra}, VLA}


\software{\textsc{aplpy} \citep{aplpy},  
          \textsc{astropy} \citep{astropy},
          \textsc{pandas} \citep{pandas},
          \textsc{matplotlib} \citep{matplotlib},
          \textsc{scipy} \citep{scipy},
          \textsc{astrodendro} \citep[\url{http://www.dendrograms.org/},][]{Rosolowsky2008}}

\bibliography{references}{}
\bibliographystyle{aasjournal}



\end{document}